\documentstyle[prl,aps,epsfig]{revtex}
\draft
\begin{document}
\twocolumn[
\hsize\textwidth\columnwidth\hsize\csname@twocolumnfalse\endcsname

\title{Weakly-coupled Hubbard chains at half-filling and Confinement}
\author{Karyn Le Hur\cite{New}}
\address{Theoretische Physik, ETH-H\"onggerberg, CH-8093 Z\"urich, Switzerland}
 \maketitle

\begin{abstract}
We study two (very) weakly-coupled Hubbard chains in the half-filled case,
and especially the situation where the intrachain Mott scale $m$ is 
much larger than the (bare) single-electron interchain hopping
$t_{\perp}$. 
First, we find that the divergence of the 
intrachain Umklapp channel at the Mott
transition results in the complete vanishing of the
single-electron interchain hopping: This is significant of a strong
confinement of coherence along the chains. 
Excitations are usual charge fermionic 
solitons and spinon-(anti)spinon pairs of the Heisenberg chain. 
Then, we show rigorously how the tunneling of spinon-(anti)spinon pairs 
produces an antiferromagnetic interchain exchange of the order of 
$J_{\perp}={t_{\perp}}^2/m$. In the ``confined'' phase and in
the far Infra Red, the system behaves as a pure spin ladder.
The final result is an insulating
ground state with spin-gapped excitations exactly as in the opposite 
``delocalized'' limit (i.e. for rather large interchain hoppings) 
where the two-leg ladder is in the well-known insulating D-Mott phase. Unlike 
for materials with an infinite number of coupled chains (Bechgaard salts), 
the confinement/deconfinement 
transition at absolute zero is here a simple crossover: no metallic phase is 
found in undoped two-leg ladders. This statement might be generalized for
N-leg ladders with N=3,4... (but not too large).

\end{abstract}
\twocolumn
\vskip.5pc ]
\narrowtext

\section{Introduction}

One-dimensional (1D) electron systems have attracted a great attention over the
last years. The Hubbard chain, that is a nice prototype to describe 1D
conductors, simply reduces to the so-called ``Luttinger liquid'' 
at sufficiently
low energy\cite{Jerome}. 
Due to the restricted motion along one direction in space,
perturbations easily propagate coherently. This induces that 
spin and charge degrees of freedom get independent, and that
all the low-energy excitations are collective modes - namely, longwavelength 
fluctuations
of the charge- or spin 
density. No Landau quasiparticle type elementary
excitations exist\cite{Haldane,Schulz}.  

Two-coupled Hubbard chains have been studied as a basic model which includes
intrachain interaction $U$, longitudinal t and transverse $t_{\perp}$ 
hoppings\cite{Articles}. For recent and general 
reviews on this subject, consult
Refs.\cite{MPA,TNG,Heinz}.
In this paper, we study the case of a weakly coupled Hubbard ladder at
(and close to) half-filling with $t_{\perp}\ll U\ll t$. It should be noted 
that this
interesting situation has been rarely investigated in the litterature.
\vskip 0.1cm
The presence of intrachain Umklapps provides 
two competing energy scales\cite{Giam1}, the Zeeman-like band
splitting energy tending to delocalize (deconfine) particles in the transverse
direction\cite{Boies}
\begin{equation}
\label{split}
\Lambda\propto t_{\perp}^{1/1-\eta}\ \hbox{with}\ \eta\propto U^2, 
\end{equation}
and the well-known single-chain Mott scale 
\begin{equation}
\label{Mottscale}
m\propto \exp-\pi v_F/U,
\end{equation}
that may rather induce confinement along the chains.
Influenced by the plethora of novel phenomena predicted (postulated) at the
confinement/deconfinement transition in systems with an infinite number
of {\it weakly} coupled chains (Bechgaard salts)\cite{Vescoli}, 
it is naturally worth to consider such issues on a two-leg
ladder where tractable calculations are indeed possible.

When the splitting energy dominates - i.e. when the hopping 
amplitude increases much 
faster than the Umklapp channel(s) upon renormalization -
that would correspond to the deconfinement phase. The result is a two-band 
model with four Fermi points.  
Rewriting the bare interactions - Do consult Eq.(\ref{int}) - in the 
so-called band basis, we
immediately recover the model that has been previously studied by Lin, 
Balents and Fisher in the limit of very large interchain hoppings 
$\Lambda\gg U$\cite{LBF}. The
renormalization group transformation scales the system to a special 
strong-coupling Hamiltonian
with enormous symmetry - the SO(8) Gross-Neveu model, that
is integrable. This allows to conclude that as soon as $\Lambda\gg m$
the two-leg ladder system is in the so-called D-Mott insulating
phase with spin-gapped excitations
and ``preformed'' short-range d-wave pairing correlations\cite{gene}. 

The splitting of the two bands affects spin-charge separation:
Cooper pairs and magnons 
appear as natural excitations both at the two-band Mott scale 
$M\approx t_{\perp}\exp-\pi v_F/8U$ that is of the same order as m. 
Doping such a spin liquid liberates ``hole-pairs'' with
d-wave pairing, restoring partially spin-charge separation\cite{MPA,Heinz}.
There is an extended massive $\pi$-mode 
with S0(6) symmetry, containing remnant charge and spin 
degrees of freedom\cite{Heinz-SO(N)}.

Then, we study this model of interacting electrons hopping on
a two-leg ladder, focusing on the behavior at half-filling mainly
in the opposite limit where the intrachain Umklapp vertex is flowing
first to strong couplings - i.e. for very small interchain hoppings. 
First, we show that 
the divergence of the Umklapp channel at the Mott transition 
makes the single-electron interchain transfer completely vanishing, 
resulting in perfect confinement of coherence along the chains.
We insist on the fact that the Kosterlitz-Thouless like
transition is then associated with a jump both in the Luttinger liquid 
exponent and in the single-electron interchain transfer. 

We may explain the entire suppression
of the latter at the Mott scale, as a consequence
of the strong breakup of physical electrons. The system can be
also seen as a superposition of two degenerate bands.
A noticeable fact is that in absence of Umklapps i.e. away from
half-filling, this condition of confinement is never satisfied for on-site
(weak) interactions\cite{Anderson}. However, this could happen
at zero temperature in presence of long-range 
interactions along the chains\cite{HeinzS,remw} 
(or still low-frequency Holstein phonons), or possibly in very 
restrictive models extended to an infinite number of coupled Hubbard 
chains\cite{KFLE}. 

At the 
opening of the Mott gap, low-lying excitations located mainly
on the chains are usual empty/doubly occupied sites and spinon-(anti)spinon 
pairs. 
Then, as a result of spin-charge separation arising in the confined
phase, we show rigorously how the tunneling of spinon-(anti)spinon 
pairs give rise to an interchain Heisenberg exchange coupling of the order of 
$J_{\perp}={t_{\perp}}^2/m>0$\cite{K-M}. This has been previously neglected in 
Ref.\cite{Suzumura}. 
The result is equivalent to two weakly 
coupled Heisenberg chains\cite{Dagotto}. 
The ground state is a disordered spin liquid with a
prominent $4k_F$ Charge Density Wave (CDW), 
like in the deconfined limit. It is worth noting that the spin gap 
$\approx J_{\perp}$ 
rejoins the charge 
gap in the entrance of the D-Mott phase. 

To summarize: The confinement/deconfinement transition arising in 
undoped two-leg ladder systems is a pure crossover\cite{Giamcom}. 
No metallic phase 
is found. This statement might be generalized for a N-leg ladder with
finite N $(N=3,4...)$.
Conclusions are really different than for an infinite number of 
coupled chains where the deconfinement transition\cite{CBourb,Giam1} 
- i.e. when formally $\Lambda\simeq m$ -
would be {\it \`a priori} responsible for 
the difference of behavior between the 
TMTTF (1D insulator) and the TMTSF (strange metal) Bechgaard 
salts\cite{Vescoli}. 

The plan of the paper is as follows.
In Sec. II, the model and formalism are
introduced. In Sec. III, based on Renormalization Group arguments and
bosonization tools, we define the confinement phenomenon. In Sec. IV, we study
in details the spinon pair hopping: The system behaves as a
pure spin ladder. In Sec. V, doping effect on the confined state 
and incoherence of $t_{\perp}$ away from half-filling are carefully
studied. In Sec. VI, we formulate a brief summary of our results
and make comparisons with Bechgaard salts.
In Appendix A, (Abelian) definitions of spinon-
and charge excitations below the Mott scale are given. In Appendix B, 
``anyonic'' charge excitations
of Luttinger liquids are reminded.

\section{Model}

We start with the following model. The kinetic
energy takes the form ${\cal H}_{kin}={\cal H}_o+{\cal H}_{\perp}$
\begin{eqnarray}
{\cal H}_o&=&-t\sum_{j,\alpha} d_{j\alpha}^{\dagger}(x+1)d_{j\alpha}(x)
    +{\mathrm H.c.}
  \nonumber\\
{\cal H}_{\perp}&=&-t_{\perp}\sum_{\alpha}d_{2\alpha}^{\dagger}(x)d_{1\alpha}(x)+{\mathrm H.c.}
\end{eqnarray}
The indices $j=1,2$ denote here the {\it chains}, and 
$\alpha=\uparrow,\downarrow$.
Focusing on electronic states near the Fermi points, one can expand:
\begin{equation}
\label{def}
d_{j\alpha}(x)=d_{+j\alpha}(x)e^{ik_Fx}+d_{-j\alpha}(x)e^{-ik_Fx}.
\end{equation}
At half-filling, one must equate $k_F=\pi/2$. We put the bare short-distance
cutoff, $a=1$. Then, the Hubbard four-Fermion interaction can be expressed
in terms of currents, defined as\cite{LBF,gene}
\begin{eqnarray}
J_{pjj}=\sum_{\alpha}d_{pj\alpha}^{\dagger}d_{pj\alpha},\hskip 0.2cm & & 
{\bf
  J}_{pjj}=\frac{1}{2}\sum_{\alpha,\alpha^{\prime}}d_{pj\alpha}^{\dagger}
{\pmb{$\sigma$}}_{\alpha\alpha^{\prime}}d_{pj\alpha^{\prime}},\\ \nonumber
& &  I_{pjj}=\sum_{\alpha,\alpha^{\prime}}d_{pj\alpha}
\epsilon_{\alpha\alpha^{\prime}}d_{pj\alpha^{\prime}},
\end{eqnarray}
and in the following $p=\pm$ denote respectively right and 
left excitations. ${\pmb{$\sigma$}}$ denote Pauli matrices and
$\epsilon_{\alpha\alpha^{\prime}}$ is antisymmetric: 
$\epsilon_{\alpha\alpha^{\prime}}=-\epsilon_{\alpha^{\prime}\alpha}$
and $\epsilon_{\uparrow\downarrow}=1$.
Precisely, the set
of marginal momentum (and non-)conserving four-Fermion interaction reads:
\begin{equation}
\label{int}
{\cal H}_{int}=\sum_j\ 
g_c J_{+jj}J_{-jj}-g_s {\mathbf J}_{+jj}{\mathbf J}_{-jj}
+g_u I_{+jj}^{\dagger}I_{-jj}.
\end{equation}
Note that
$g_c$ and $g_s$ describe charge- and spin backscatterings respectively,
and $g_u$ the intrachain umklapp processes. The bare interactions are of the
order of $U$.

Now, it is appropriate to use an Abelian bosonized form for 
fermion
operators. These transform as \cite{Haldane,Schulz}
\begin{equation}
\label{bos}
d_{pj\alpha}=\exp\hbox{\Large(}i\sqrt{\frac{\pi}{2}}\left[p(\Phi_{jc}+\alpha\Phi_{js})
    -(\theta_{jc}+\alpha\theta_{js})\right]\hbox{\Large )}.
\end{equation}
$\alpha=\pm$ for spin up and spin down, respectively.
Remarkably, electron spectrum yields spin-charge
separation. Absorbing the interaction $g_c$, that is not affected
by a rescaling of the short-distance cutoff, in the charge part
of ${\cal H}_o$ results in the so-called Luttinger model
\begin{equation}
H_{oj}^c=\frac{u}{2\pi}\int dx\ \frac{1}{K}{(\rho_{jc}-\rho_o)}^2
+K{\nabla\Theta_{jc}}^2.
\end{equation}
$\partial_x\Phi_{jc}=(\rho_{jc}-\rho_o)$ measures
fluctuations of charge density in each chain, and $\nabla\Theta_{jc}$ is the
conjugate momentum to $\Phi_{jc}$. All the interaction effects are now hidden
in the parameters u (the velocity of charge excitations, $uK=v_F$ with
the Fermi velocity $v_F=2t\sin k_F$) and
K (the Luttinger liquid (LL) exponent controlling the decay of 
correlation functions, $K=1-g_c/\pi v_F$). $K<1$ means repulsive
interactions. 

The free spin Hamiltonian yields the same form replacing
$\Phi_{jc}$ by $\Phi_{js}$ and the Luttinger exponent K by $K_s=1$ due to the
requirement of SU(2) invariance.  

\section{Confinement due to Umklapps}

It is first suitable to write the main Renormalization Group (RG)
equations.

\subsection{RG-flow analysis}

By simple scaling arguments, one gets:
\begin{eqnarray}
\label{flow}
\frac{dg_u}{dl} &=& (2-2K)g_u,\\ \nonumber
 \frac{d\ln t_{\perp}}{dl} &=& \frac{3}{2}-\frac{1}{4}(K+\frac{1}{K}).
\end{eqnarray}
Anisotropy between space and time induced by Umklapp scattering results in a
strong renormalization of the LL exponent and the charge velocity\cite{Giam2}:
\begin{eqnarray}
\label{flow2}
\frac{dK}{dl} &=& -C_1 (g_uK)^2,\\ \nonumber
\frac{du}{dl} &=& -C_2{g_u}^2uK.
\end{eqnarray}
For a complete derivation of such equations, do consult Ref.\cite{TH}.
The $l$ describes renormalization of the short-distance cutoff $a(l)=\exp l$.
At half-filling, usual Bessel functions 
$J_0$ and $J_2$ are non-oscillating and have been included
in the positive constants $C_1$ and $C_2$\cite{Giam2}. If one is at finite
temperature T, then the renormalization procedure must be stopped at lengths
$a(l)$ comparable to the thermal length u/T, that means $l\sim \ln(1/T)$.

Starting with a 
sufficiently small (bare) interchain hopping, then the Umklapp channel
$g_u$ is flowing first to strong couplings at a critical scale $T_c$ 
that is the
single-chain Mott scale m given by Eq.(\ref{Mottscale}).
For one-loop RG equations, indeed one gets a
critical driving parameter $l_c=\alpha v_F/U$ where $\alpha$ is of
the order of unity. 
Although the singularity at finite $l_c$ is an artifact
of the one-loop calculations, which is actually eliminated by including
higher-order terms, m always remains a meaningful characteristic energy
scale for strong coupling (see Appendix A). This is significant of a Mott 
{\it insulating} state, with a finite charge gap equal to 2m.

An important point that can be extracted from the RG-flow is that the explicit
divergence of the Umklapp scattering $g_u$ at the opening of the Mott gap
results formally in:
\begin{equation}
\label{K}
K(l>l_c)=0. 
\end{equation}
The jump of the LL exponent
at finite T is significant of a Kosterlitz-Thouless transition, that is
also accompanied by a sharp jump in the charge compressibility and Drude
weight\cite{Giam1}. 
Likewise, we like to stress on the fact that the strong 
decrease of the LL
exponent by Umklapps has an enormous consequence on the small single-particle
interchain hopping as well. 

Consulting Eqs.(\ref{flow}),(\ref{K}), indeed we observe 
that this
should also produce the complete vanishing of $t_{\perp}$ exactly at the
Mott scale (and not only a partial reduction\cite{Kishine}):
\begin{equation}
t_{\perp}(l>l_c)=0.
\end{equation}
In agreement with Giamarchi's assertion, we obtain that the Mott gap renders
the single-particle hopping completely irrelevant\cite{Giam1}. We like
to interpret
the total vanishing of $t_{\perp}$ at finite temperature as a 
sign of strong confinement (of coherence) along the chains. For a 
physical explanation, do see next subsection. 
The divergence of intrachain
Umklapp scattering produces in that case an insulating state with two 
degenerate bands.
\vskip 0.1cm
Now, we like to repeat that in absence of Umklapp scattering i.e. away from
half-filling, this condition of confinement is never satisfied for the 2-leg
Hubbard ladder\cite{Anderson}. Precisely, to make interchain hopping 
irrelevant away from
half-filling, one should make its amplitude decreasing upon renormalization.
This must be determined from inequality $d_{\perp}>2$ with 
\begin{equation}
d_{\perp}=\frac{1}{4}(K+\frac{1}{K})+\frac{1}{2},
\end{equation}
the scaling dimension of the $t_{\perp}$-perturbation. To obey this standard 
criterion of
irrelevance\cite{TNG}, one should start with a very
small bare LL exponent $K<3-\sqrt{8}\simeq 0.171$...: It is
worth to note that this is never  
realized for purely on-site interactions and $U\ll t$\cite{HeinzS}. 
This always produces deconfinement away from half-filling 
that is characterized by a finite Zeeman-like splitting of degenerate bands. 
The magnitude of the splitting $\Lambda$ is defined in 
Eq.(\ref{split}). The system behaves as a Luther-Emery liquid\cite{LE} with
dominant d-wave pairing.
For a brief description of the corresponding phase, see
Section V. 
\vskip 0.1cm
Nonetheless, the confinement condition away from half-filling 
might be performed in presence of long-range
interactions where the chains behave rather as Wigner crystals
\cite{HeinzS}. The underlying model in each chain 
is still of LL type, but the system
develops a real
tendency to a periodic arrangement or a $4k_F$ CDW: This hinders
considerably 
the single-particle interchain transfer. It should be remarked that in that
case, confinement might take place only at zero temperature.

We also like to insist on the following point.
Starting with very small interchain hoppings (such as $m\gg t_{\perp}$) 
ensures $g_s^*=0$ at zero temperature: Indeed, for the 1D
Hubbard chain and repulsive interactions it is well-known that the spin 
backscattering is (marginally) 
irrelevant and then it can be omitted\cite{Schulz}. 
Note already the fundamental difference with
the deconfined regime, where spin backscattering also flows to strong
couplings at the D-Mott transition producing immediately spin-gapped
excitations\cite{MPA,LBF,remark}.

\subsection{Breakup of electrons on the chains}

The impossibility to get a finite
single-particle interchain transfer at absolute zero 
might be naturally interpreted as follows. The system
suppresses the bare-electron interchain hopping due to the strong constraint
to avoid a double on-site occupancy, that is here explicitly induced by
the presence of the finite Mott gap. 
Now, we really like to 
emphasize that the vanishing of $t_{\perp}$ 
{\it immediately at the Mott transition}
``caches'' more 
a marked spin-charge separation or the complete breakup of physical
electrons. 

When $T\rightarrow m$, 
one already finds that the electronic Green function (e.g for Right movers)
\begin{equation}
{\cal
G}_+(x,\tau)=\frac{e^{ik_Fx}}{\sqrt{(v_F\tau-ix)(u\tau-ix)}}[x^2+(u\tau)^2]^{-\eta/2},
\end{equation} 
tends to vanish for quite short time because
the anomalous dimension of the electron increases fastly:
$\eta=1/4(K+K^{-1})-1/2\gg 2$ through the factor
$K^{-1}(l_c)$. As usual, $\tau$ denotes the Matsubara
Imaginary time. 
Now, we show precisely that {\it below} the Mott transition there is no way to
recombine a physical electron along the chains: This gives a simple 
explanation on the total vanishing of the bare-electron interchain transfer 
for $T\leq m$. 

First, from Eqs.(\ref{flow2}), one can observe that 
the velocity of charge excitations drastically decreases to zero. 
This tends to push forward the idea that charge and spin degrees
of freedom get really independent. Precisely, the opening of the Mott gap 
transforms charge excitations on the chains into
{\it fermionic Kinks}, corresponding
to pairs of doubly and empty occupied sites. The spin sector is still
described by {\it spinon-(anti)spinon pairs} of the Heisenberg model. 
Do consult Appendix A for more explanations and details. 

In the following, the symbols ${\cal F}_{pj}^{Q_c^\pm}$ and 
${\cal S}^{Q_s^\pm}_{pj}$ refer to 
a fermionic Kink and a (single) spinon, respectively. These are given 
by: 
\begin{eqnarray}
{\cal F}^{Q_c^\pm}_{pj}&\approx &\exp i\sqrt{\pi}Q_c^{\pm}(-p\tilde{\Phi}_{jc}+\tilde{\Theta}_{jc}),\\ \nonumber
{\cal S}^{Q_s^\pm}_{pj}&=&\exp i\sqrt{\frac{\pi}{2}}Q_s^{\pm}(-p{\Phi}_{js}+
\Theta_{js}).
\end{eqnarray}
Note that $Q_c^{\pm}=\pm$ refer to ``electron'' and ``hole'' like excitations
precisely, and the spin of the spinon obeys $S^z=Q_s^{\pm}/2=\pm 1/2$. 
The spinon
and charge objects have different wave-vectors; an empty
or doubly occupied site has a (double and opposite) wave-vector 
$(-pQ_c^{\pm})2k_F$.

Second using Eq.(\ref{bos}), it is advantageous to decompose the bare 
electron operator
\begin{equation}
\label{electron}
d^{\dagger}_{pj\alpha}=\hbox{\Large{[}}{\cal C}^{+}_{pj}{\cal S}^{\pm}_{pj}
\hbox{\Large{]}},
\end{equation}
where ${\cal C}^{+}_{pj}$ describes the holon of the Fermi gas\cite{K.-V}:
\begin{equation}
{\cal C}^{Q_c^\pm}_{pj}=\exp i\sqrt{\frac{\pi}{2}}Q_c^{\pm}(-p{\Phi}_{jc}+
\Theta_{jc}).
\end{equation}

Now, it is sufficient to note that starting with a bare LL exponent 
$K\rightarrow 1$ (i.e weak U), one can equate:
\begin{equation}
{\cal C}^{+}_{pj}=\hbox{\Large{[}}{\cal F}^{+}_{pj}\hbox{\Large{]}}
^{1/\sqrt{2}}.
\end{equation}
This may simply explain why the recombination of
electrons is definitely forbidden at the Mott transition: 
{\it Indeed, this would
require a fractional number of fermionic Kinks}.
As a natural consequence, the single-particle interchain
transfer is already destroyed by the enhanced spin-charge
separation arising at the Mott transition. This constitutes the main difference
with the confinement phenomenon in Wigner crystals that occurs only at
absolute zero.

\subsection{Prevalent fluctuations at the Mott transition}

Using known results on the Hubbard chain at half-filling
and those of Appendix A, we get that the $4k_F$ CDW and spin-spin 
fluctuations at $q=\pi$ are then prominent around the Mott scale
with the most diverging susceptibility. 

Here, these are described by the operator:
\begin{equation}
O_{cdw}^j={\cal F}_{+j}^{+}{\cal F}_{-j}^{-}+{\mathrm H.c.},
\end{equation}
that yields a non-zero expectation value below the Mott transition, 
resulting explicitly in:
\begin{equation}
{<O_{cdw}^j(x)O_{cdw}^j(0)>}={<O_{cdw}^j(x)>}^2=const.,
\end{equation}
and by the staggered magnetization operator: 
\begin{equation}
{\bf m}_j={\cal
  S}_{-j}^{\mu}{\pmb{$\sigma$}}({\cal
  S}_{+j}^{\mu^{\prime}})^*+{\mathrm H.c.}
\end{equation}
The spinon operator has the quantum scaling dimension 1/2, that produces:
\begin{equation}
<{\bf m}_j(x){\bf m}_j(0)>\ =(-1)^x/x.
\end{equation}
Since we have a jump in the
LL exponent at the Mott transition, we recover Heisenberg correlation
functions in each chain.
\vskip 0.03cm
Again unlike in the deconfined picture\cite{MPA,LBF}, here 
the opening of the Mott gap 
does not produce any spin gap. The single chain behaves as a pure Heisenberg
spin chain at the Mott transition. Our picture below the Mott energy, however, is still incomplete. The tendency
towards suppression of single-particle transport in the transverse 
direction
is not the only effect of $t_{\perp}$\cite{CBourb}. 
The Hamiltonian must be supplemented
by extra ``relevant'' 
terms which are inevitably generated in the course of
renormalization, by expanding the partition function
as a function of $t_{\perp}$\cite{K-M}. For a detailed review on this
important point, do see Ref.\cite{TNG}.

\section{Tunneling process}

It is maybe appropriate to rewrite explicitly the bare forward and backward
hoppings\cite{K-M,rem}:
\begin{eqnarray}
{\cal H}_{\perp}&=&
  t_{\perp}\cos\sqrt{\pi}\Theta_c^-\cos\sqrt{\pi}\Theta_s^-
\hbox{\Large{[}}\cos\sqrt{\pi}\Phi_c^-\cos\sqrt{\pi}\Phi_s^-\\ \nonumber
&+&(-1)^x\cos\sqrt{\pi}\Phi_c^+\cos\sqrt{\pi}\Phi_s^+\hbox{\Large{]}},
\end{eqnarray}
combining the boson fields in the two chains into a symmetric ``+'' 
and antisymmetric ``-'' part. On the other hand, using Eqs.(\ref{def})
and (\ref{electron}) one can also write:
\begin{eqnarray}
\label{hop}
{\cal H}_{\perp}=\sum_{p,p^{\prime}}\ &t_{\perp}&e^{i(p^{\prime}-p)k_Fx}
\hbox{\Large{[}}{\cal C}^{+}_{p1}{\cal C}^{-}_{p^{\prime}2}+
{\mathrm H.c.}\hbox{\Large{]}}\\ \nonumber
&\times&\hbox{\Large{[}}{\cal S}^{+}_{p1}{\cal S}^{-}_{p^{\prime}2}+
{\mathrm H.c.}\hbox{\Large{]}}.
\end{eqnarray}
Indeed, one always may interpret 
the hopping of a bare electron from one
chain to the other as the hopping of a pure semionic charge
$Q_c^+=+1$ and a single spinon with spin $S^z=\pm 1/2$: The two objects 
behave then as half-electrons. 

As a consequence, we get the important identifications:
\begin{eqnarray}
\label{tt}
{\cal C}^{+}_{p1}{\cal C}^{-}_{p2} +{\mathrm H.c.}&=&
\cos\sqrt{\pi}\Theta_c^-\cos\sqrt{\pi}\Phi_c^-,\\ \nonumber
{\cal C}^{+}_{p1}{\cal C}^{-}_{-p2}+{\mathrm H.c.} &=&
\cos\sqrt{\pi}\Theta_c^-\cos\sqrt{\pi}\Phi_c^+,
\end{eqnarray}
and similarly for the spinon part.

\subsection{Spinon pair hopping}

As shown before, the total vanishing of $t_{\perp}$ for $T\ll m$ must be 
closely related to the breakup of electrons on the chains. On the other hand,
we insist on the fact that ``some'' tunneling processes in 
${t_{\perp}}^2$ -
respecting spinon and charge fermionic Kink separation - are still allowed in 
the far Infra-Red for $T\leq m$. Here, we like to 
emphasize that in previous works
spin excitations at very low temperature have not been analyzed in great 
details\cite{K-M}, or relevant tunneling
processes have been completely undervalued\cite{Suzumura}. 
\vskip 0.1cm
Now, we derive these properly as follows. Note that below
we implicitly drop out
oscillatory terms that do not influence the physics at the fixed point.

First, it is important to mention that at the Mott transition 
$(l\rightarrow l_c)$, the relevant 
contributions are essentially furnished by:
\begin{eqnarray}
\hbox{\Large{(}}{\cal C}^{+}_{p1}{\cal C}^{-}_{p2}+{\mathrm H.c.}
\hbox{\Large{)}}\hbox{\Large{(}}
{\cal C}^{+}_{p^{''}1}{\cal C}^{-}_{p^{''}2}+{\mathrm H.c.}
\hbox{\Large{)}}
&=&\cos\sqrt{4\pi}\Phi_c^-,\\ \nonumber
\hbox{\Large{(}}{\cal C}^{+}_{p1}{\cal C}^{-}_{-p2}+{\mathrm H.c.}
\hbox{\Large{)}}\hbox{\Large{(}}
{\cal C}^{+}_{p^{''}1}{\cal C}^{-}_{-p^{''}2}+{\mathrm H.c.}
\hbox{\Large{)}}
&=&\cos\sqrt{4\pi}\Phi_c^+.
\end{eqnarray}
Such operators acquire indeed a non-zero expectation value (see Appendix A):
\begin{equation}
\label{cut}
<\cos\sqrt{4\pi}\Phi_c^->\ =\ <\cos\sqrt{4\pi}\Phi_c^+>\ \approx m,
\end{equation}
whereas one still gets:
\begin{equation}
<\cos\sqrt{4\pi}\Theta_c^->\ =0.
\end{equation}
These may be primarily interpreted as tunneling
processes of the massive fermionic Kinks lying along the chains. For 
a bare LL exponent $K\rightarrow 1$ one precisely gets:
\begin{eqnarray} 
\hbox{\Large{[}}({\cal F}^{+}_{p1}{\cal F}^{-}_{p^{\prime}2})+{\mathrm H.c.}\hbox{\Large{]}}^2 &=& \hbox{\Large{[}}({\cal C}^{+}_{p1}{\cal C}^{-}_{p^{\prime}2})+{\mathrm H.c.}\hbox{\Large{]}}^{2\sqrt{2}},\\ \nonumber
&=& m\hbox{\Large{[}}({\cal C}^{+}_{p1}{\cal C}^{-}_{p^{\prime}2})+
{\mathrm H.c.}\hbox{\Large{]}}^2.
\end{eqnarray}
Approaching the
Mott transition, second 
it is important to note that the (finite) amplitude of the relevant
pair hopping(s) is given approximately by [See Appendix A and Eq. (A20)]:
\begin{equation}
[t_{\perp}(l\rightarrow l_c)]^2
/{E_F}^2\approx {t_{\perp}}^2\exp 2l_c/{E_F}^2={t_{\perp}}^2
m^{-2}. 
\end{equation}
$E_F$ is the Fermi energy.
Now, averaging 
on the charge sector for $l\approx l_c$, one gets the extra relevant 
Hamiltonian:
\begin{equation}
\delta{\cal H}_{int}={t_{\perp}}^2/m
\hbox{\Large{[}}\sum_{p,p^{\prime}} 
{\cal S}^{+}_{p1}{\cal S}^{-}_{p^{\prime}2}+
{\mathrm H.c.}\hbox{\Large{]}}^2.
\end{equation}
$\delta{\cal H}_{int}$ produces processes of coherent interchain
spinon-(anti)spinon hopping triggered by the single particle one.

Using the classification scheme of
particle-particle and particle-hole hoppings from
Ref.\cite{K-M}, one can equivalently rewrite\cite{rem0}:
\begin{eqnarray}
\delta{\cal
  H}_{int}=&g_5&\cos\sqrt{4\pi}\Phi_s^-+(g_1+g_4)\cos\sqrt{4\pi}\Theta_s^-
\\ \nonumber
&+&g_8\cos\sqrt{4\pi}\Phi_s^+.
\end{eqnarray}
where we have: $g_i={t_{\perp}}^2/m$. Moreover, from Appendix A, it is 
then clear that this generates an
interchain Heisenberg interaction:
\begin{equation}
\delta{\cal H}_{int}=J_{\perp}\ {\bf m}_1\cdot{\bf m}_2,\qquad
J_{\perp}={t_{\perp}}^2/m\ll m.
\end{equation}
Since the spinon spectrum (driven by the kinetic part)
is known to be invariant tuning the Hubbard interaction from
weak to strong interactions, therefore the low-energy model is equivalent to
{\it two weakly-coupled Heisenberg chains}\cite{Dagotto}. 
The intrachain Heisenberg coupling is
here equal to the Fermi velocity, $v_F$. 

This proves rigorously that the
interchain hopping is well sufficient to generate the antiferromagnetic spin
exchange $J_{\perp}\approx {t_{\perp}}^2/m>0$, that will make all spin
excitations gapful. In particular, semi-classical considerations allow to
predict the pinning of the $\Theta_s^-$ and $\Phi_s^+$ spin-fields at the
fixed point (in the minima of the cosine potentials). Excitations will be
separated from the ground state by an energy gap $\sim J_{\perp}$. 
This leads to a ground state that would {\it not} have still a strong 
resemblance
to that of the half-filled 1D chain. Magnetic excitations and doping effects
are indeed very different.

\subsection{Effective Heisenberg ladder}

Excitations of the final spin Hamiltonian $(H_o^s+\delta{H}_{int})$ have been 
studied in
great details in Refs.\cite{TNG,Tsvelik}. Rather than belabor the 
derivation, we only 
indicate the results. Rewriting the model in terms of 
four (real) Majorana fermions ${\pmb{$\xi$}}=\sum_{i=1}^3\xi_i$ and $\rho$, one gets the
following conclusions.

As for a general S=1 magnet, the triplet excitations in a rung
namely ${\pmb{$\xi$}}$ will have a gap - the
Haldane gap\cite{Haldane2} - $m_t=J_{\perp}$.
The singlet branch $\rho$ is located at quite high energy, of the order of
$m_s=-3J_{\perp}$\cite{Tsvelik}. For $T\ll J_{\perp}$, 
the asymptotic
correlation function
for spins on the same chain is given mainly by
\begin{equation}
<{\bf m}_j(x){\bf m}_j(0)>\ = (-1)^x{x}^{-1/2}\exp(-m_t x).
\end{equation}
This definitely produces
 a short-range Resonating Valence Bond solid exactly as in the 
deconfined limit.

Note that taking only into account the (bare) interchain
hopping term at a short wave vector $q=0$, 
some spin excitations would remain
gapless at the Infra-Red fixed point:
\begin{equation}
\delta{\cal H}_{int}=g_5\cos\sqrt{4\pi}\Phi_s^-
+g_1\cos\sqrt{4\pi}\Theta_s^-.
\end{equation}
The $(s^-)$ modes are protected by the duality symmetry under
$\Phi_s^-\leftrightarrow \Theta_s^-$, the $(s^-)$ sector would be in fact a 
critical
point of the two-dimensional Ising type.
The resulting Hamiltonian must
 be assigned to the same universality class of
the purely {\it forward} scattering model considered by Finkel'stein and 
Larkin\cite{F-L} and later
by Schulz\cite{Heinz}, resulting in 
gapless- spinons and singlet (Majorana) fermions. These excitations can be
recombined and rewritten as a massless triplet magnon 
excitation ${\pmb{$\xi$}}$ in a rung. 

But, in the two-chain problem away from half-filling, again
it is known that for {\it general} repulsive interactions, there is always a 
spin excitation gap and d-type pairing - or exceptionally orbital 
antiferromagnetism fluctuations. 
Here, the presence of such enigmatic spin liquid
with remnant gapless magnon modes is definitely forbidden by the bare
backward hopping at large wave vector $q=\pi$. 

To summarize, we really want to emphasize that unlike in the deconfined
region, the latter plays a central role in presence of strong confinement
along the chains.

\subsection{Crossover to the D-Mott state}

We can now conclude that
the two weakly-coupled Hubbard chain model is in the same phase C0S0 for
very {\it small} bare interchain hoppings (confined region) and for rather
{\it large} bare interchain hoppings (deconfined phase); 
CnSv denotes a state with n massless charge and v massless spin modes.
This naturally demonstrates that the
confinement/deconfinement transition at absolute zero 
in the half-filled two-chain problem is a simple crossover\cite{Giamcom}. 
Now, let us compare symmetries of excited states in these two regimes.

In the confined regime, we have a total spin-charge separation 
on the {\it chains}. Charge
and spin excitations are gapped but ruled by different energy scales. 
Additionally, we have shown that far below the Mott transition excitations
can be classified as charge fermionic Kinks lying mainly along
the chains with symmetry U(1) and charge $\pm 1$, and magnon
like excitations with an underlying symmetry 
$SU(2)_2\times{\cal Z}_2$\cite{symmetry}.

In the deconfined regime, the effective model yields rather
an enormous global S0(8) symmetry that can be briefly 
understood in the {\it band} basis, as follows\cite{MPA,LBF}. The low-energy 
physics depends on
a single effective coupling constant g (of the order of U). In the
band picture, the interaction part takes the specific form:
\begin{equation}
{\cal H}_{int}=-g\cos\sqrt{4\pi}\Phi_a\cos\sqrt{4\pi}\Phi_b,
\end{equation}
with $(a,b)=1,2,3,4$ and bosonic operators are given by:
\begin{eqnarray}
(\Phi,\Theta)_1 &=& (\Phi,\Theta)_{\rho}^+\qquad (\Phi,\Theta)_3=(\Phi,\Theta)_{\sigma}^+,\\ \nonumber
(\Phi,\Theta)_2 &=& (\Phi,\Theta)_{\sigma}^-\qquad (\Phi,\Theta)_4=(\Theta,\Phi)_{\rho}^-.
\end{eqnarray}
The indices $_{\rho}^{\pm}$ and $_{\sigma}^{\pm}$ refer to 
symmetric/antisymmetric charge and spin fluctuations in the band basis.
It is then appropriate to use the re-Fermionization procedure 
(again, $Q_c^-=-$):
\begin{equation}
\Psi_{pa}\approx\exp i\sqrt{\pi}Q_c^-(-p\Phi_a+\Theta_a).
\end{equation}
Then, for weak U the result is\cite{MPA,LBF}:
\begin{equation}
{\cal H}=\sum_{a=1}^4\Psi^{\dagger}_a iv_F\tau^z\partial_x\Psi_a-g\hbox{\Large{(}}\Psi^{\dagger}_a\tau^y
\Psi_a\hbox{\Large{)}}^2.
\end{equation}
Pauli matrices ${\pmb{$\tau$}}$ act on right and left sectors and
$\Psi_a=(\Psi_{+a},\Psi_{-a})$. This is known 
as the S0(8) Gross-Neveu model. The latter
has a remarkable property of ``triality''
that is useful to equate various excited states in the deconfined phase. 
A remarkable fact that can be shown from integrability of the model is
that lowest excited states are magnon like excitations with spin-1, and
spinless charge $\pm 2$ or Cooperon and hole-pair. They are associated 
with {\it two} Dirac
fermion excitations. This beautifully demonstrates 
preformed pairing with an approximate d-wave symmetry 
(precisely, with a relative sign change between bonding- and
antibonding pairs) in the deconfined state. 
Unlike in the confined region, spin-charge separation is here violated.  

Finally, note that in the crossover region i.e. when $\Lambda\approx m$
the spin gap $J_{\perp}$ tends to rejoin the charge gap m.
Both become of the same order as $M\propto\exp-\pi v_F/8U$, the unique energy 
scale in the delocalized D-Mott phase. From the above analysis, one can
furthermore predict that thermally 
activated fermionic Kinks on the chains with charge
$\pm 1$ would then turn into excited Cooper pairs (in a rung).

\section{Doping effects}

Doping the D-Mott state, it is well-known that the low energy
excitations are only in the charge sector and correspond to a sound mode
of the ``hole pairs''. The phase is C1SO, 
i.e. a Luther-Emery liquid\cite{LE}. The resulting system behaves
then as a fluid of
hard-core bosons with quite long-range d-wave pairing\cite{MPA,Heinz,LBF}.

Very close to half-filling, 
note that the SO(8) symmetry still remains\cite{LBF}, but 
this gets easily broken down away from half-filling in $S0(6)\times U(1)$. 
The U(1) symmetry comes from the 
hole-pair bosonic field $\Phi_{\rho}^+$ that becomes critical, 
whereas the SO(6) symmetry results 
from the residual massive part. This is explicitly composed
of the 3 massive Dirac fermions $\Psi_2$, 
$\Psi_3$ and $\Psi_4$ that are still coupled via a Gross-Neveu interaction.
The most direct consequence is that spin-charge separation is only partially
restored: There is a sixfold degenerate {\it extended
$\pi$-mode} in the d-wave superconducting state that contains both 
massive spin and charge excitations\cite{Heinz-SO(N)}. Note that unlike
for the doped Hubbard chain, charge fluctuations in the superconducting
state can only subsist at $4k_F$\cite{LSM}.

It is important to remind that away from half-filling i.e
in absence of Umklapp scattering, the 2-leg Hubbard
ladder is therefore described by a C1SO phase with prominent d-wave pairing 
and unconventional $S0(6)\times U(1)$ symmetry.

It is noteworthy that the hole-pairing
phenomenon should subsist in strong magnetic field; The resulting 
2-band model (of spinless fermions) predicts indeed prominent p-wave
superconductivity for quite large $t_{\perp}$\cite{Urs}.

\subsection{Hole doping effect on the confined state}

For completeness, now 
we study light doping effect on the confined state properties. 
\vskip 0.22cm
We remind that the Umklapp interaction $g_u$ has still to be
taken into account for low dopings, $\delta<m$. It is therefore
advantageous to take the half-filled picture that has been derived before, 
as a natural starting point. 

As usual, we model the doping by adding a chemical potential 
$-\mu\hat{Q}_d$ with
the charge operator:
\begin{eqnarray}
\hat{Q}_d &=& \sqrt{\frac{2}{\pi}}\int dx\ \partial_x(\Phi_{1c}+\Phi_{2c}) 
\\ \nonumber
&=& 2\sqrt{\frac{K}{\pi}}\int dx\ \partial_x\tilde{\Phi}_c^+.
\end{eqnarray}
Each chain is supposed to be equally doped.
Note that the spin spectrum should not be affected by the 
low doping effect producing inevitably 
the pinning of the fields $\Phi_s^+$ and $\Theta_s^-$.
The resulting model then reads:
\begin{eqnarray}
H_c^+ &=& \frac{u}{2\pi}\int dx\ \hbox{\Large{\{}}
({\partial_x\tilde{\Phi}_c^+})^2
+({\partial_x\tilde{\Theta}_{c}^+})^2              \\ \nonumber
 &-&2m\cos(\sqrt{4\pi}\tilde{\Phi}_c^+)
-2\mu\sqrt{\frac{K}{\pi}}\partial_x\tilde{\Phi}_c^+\hbox{\Large{\}}}.
\end{eqnarray}
This Hamiltonian, describing a (purely 1D-quantum) commensurate-incommensurate
transition, has been actively studied in the litterature. For
a review, consult p. 172
of Ref.\cite{TNG}. Formally, solutions of the equations of motion can be still
written as fermionic solitons:
\begin{equation}
{\cal F}^-_{p+}=\exp i\sqrt{\pi}(p\tilde{\Phi}_c^+ -\tilde{\Theta}_{c}^+).
\end{equation}
The bottom of this band is typically at energy-scales close to $\mu_c=-m$. 
Using Jordan-Wigner duality, one can also associate the hard-core boson
field\cite{MPA}:
\begin{equation}
\Delta\approx\exp -i\sqrt{\pi}\tilde{\Theta}_{c}^+.
\end{equation}
Using the fact that the fields $\Phi_s^+$ and $\Theta_s^-$ are still 
pinned, then one can easily check that such object describes well a 
{\it hole-pair} with charge $Q_d=-2$ and zero momentum 
(Consult Ref.\cite{TNG} p. 280):
\begin{equation}
\Delta=d_{+1\uparrow}d_{-2\downarrow}\pm d_{+2\uparrow}d_{-1\downarrow}.
\end{equation} 
Since the antisymmetric charge sector is not affected by the 
chemical potential, fluctuations of charge density in
each chain become now strongly correlated:
\begin{equation}
<\partial_x\tilde{\Phi}_{1c}>\ =\ <\partial_x\tilde{\Phi}_{2c}>,
\end{equation}
inducing pairing between holes of the two chains. Physical excitations are
rather described by the pairing field $\Delta$ that carries zero
momentum (instead of ${\cal F}^-_{p+}$). We like to 
emphasize that in this case, charge objects at- and close to half-filling 
are then
different (fermionic Kinks with $Q_c^{\pm}=\pm$ turn into hole-pairs).

Furthermore,
one gets the important equality:
\begin{equation}
[\hat{Q}_d,\Delta]=-2\sqrt{K}\Delta.
\end{equation}
On the other hand, the (vertex) chiral operators ${\cal F}^-_{p+}$
or $\Delta$ must also satisfy
\cite{K.-V}:
 \begin{equation}
[\hat{Q}_d,\Delta]=Q_d\Delta=-2\Delta.
\end{equation}
Therefore, at the commensurate-incommensurate transition one checks that the
LL parameter $K\rightarrow 1$, exactly like for the lightly 
doped D-Mott state\cite{Saleur}. This reinforces the idea that 
at filling, there is no real difference between the confined- and 
deconfined phases (T=0): They react similarly by doping.
Such a universal value of the
LL exponent in the two-chain problem has been first conjectured in
Ref.\cite{Schulz-RC}, and recently reached numerically in the strong U-limit
with Density Matrix Renormalization Group approach\cite{DMRG_K} (where
the confined picture can be extended). The limit $K\approx 1$ is 
consistent with the picture of a very dilute hole-pair gas. More generally, 
this describes well a hard-core boson gas in low-density limit\cite{Hal}.
Note the difference with the one-chain case\cite{MI-tran}.

One can also understand the preceding result
as follows. The holes stay in the same rung because they do not break spin
singlets. The total charge mode $\Phi_c^+$ becomes massless\cite{K-M} 
resulting in an expected Luther-Emery liquid (C1S0)\cite{LE}.
Similarly to the lightly 
doped- Heisenberg ladder\cite{MGS,TNG} or D-Mott state\cite{MPA,Heinz}, 
one immediately recovers prominent d-wave superconductivity with the
pairing correlation functions:
\begin{equation}
<{\Delta}^{+}(x)\Delta(0)>\ \approx x^{-1/2}.
\end{equation}
Note that adding eventually a tiny Coulomb 
repulsion interaction $\tilde{u}$ between the chains, the triplet mass 
evolves slightly as $m_t=J_{\perp}-\tilde{u}$, but remains finite\cite{TNG}.

Finally, for larger doping i.e when $\delta>m$, the intrachain Umklapp 
channel can be neglected resulting in total deconfinement, and the phase 
C1SO acquires an enlarged $S0(6)\times U(1)$ symmetry and a sixfold 
degenerate ``$\pi$-mode'' as well. Note that the hole-pair mode 
can be
equally described in terms of the ${\Theta}_c^+$ or ${\Theta}_{\rho}^+$ phases.

\subsection{Incoherence of $t_{\perp}$ away from haf-filling}

Finally, we would like to discuss the fact that the Luttinger model
is {\it not} a good fixed point of the 2-leg Hubbard ladder
away from half-filling. This can be  
interpreted as an incoherent effect of $t_{\perp}$ because
the consequent renormalization of the interchain hopping 
away from half-filling turns the 
weak-coupling LL behavior onto a strong-coupling Luther-Emery fixed point.
We may understand the incoherence of $t_{\perp}$ mathematically, as follows.
An alternative definition of incoherence of the single-particle interchain 
transfer in LL's has been given in Ref.\cite{Clarke-Strong}.
\vskip 0.1cm
For instance, one can check that
charge eigenstates of the Luttinger Hamiltonian\cite{K.-V} 
(Consult Appendix B)
\begin{equation}
{\cal L}^{\pm}_{pj}=\exp i\sqrt{\frac{\pi}{2}}Q_c^{\pm}
(-p\frac{{\Phi}_{jc}}{K}+\Theta_{jc}),
\end{equation}
are not (exactly) equal to ${\cal C}^{\pm}_{pj}$.
As a result, the consequent renormalization of $t_{\perp}$ progressively
hinders the coherence of anyonic-type excitations lying along the 
chains. The resulting competition between the interchain hopping
$t_{\perp}$ and the (bare) interaction U
produces inevitably a strong coupling fixed point with 
no charge object with charge $Q_c^{\pm}=\pm 1$ (but paired holes) and then no
spinon (but magnons).

More generally, for doped N-leg Hubbard ladders $(N\geq 2)$
this also results in a complicated strong coupling
fixed point with new stable excitations\cite{N-chain}. 
A LL fixed point, however, can still arise in some 
{\it specific} models e.g in the 3-leg ladder (or
in the N-leg ladder, with N odd) very close to half-filling\cite{gene} or 
for a very anisotropic and restrictive 
network built with an infinite number of ``very strongly''
coupled chains\cite{KFLE}.

Finally, $t_{\perp}$ gets coherent only in the non-interacting case $(K=1)$. 
Spin and charge excitations
have the same velocity and then eigenstates in each chain
are usual Landau particles:
\begin{equation}
d^{\dagger}_{pj\alpha}=\hbox{\Large{[}}{\cal L}^{+}_{pj}{\cal S}^{\pm}_{pj}
\hbox{\Large{]}}_{K=1}.
\end{equation}
The ground state is simply achieved by making symmetric and antisymmetric
combinations of the particle operators in the two gas. The fixed
point is still a Fermi gas.

\section{Conclusion and Discussion}

To summarize briefly, we have shown that the weakly-coupled two Hubbard
chain model is in the same phase C0S0, 
for {\it irrelevant} interchain hoppings (confined region with
$SU(2)_2\times{\cal Z}_2\times[U(1)]^2$ symmetry) and for {\it relevant}
interchain hoppings (deconfined region with S0(8) symmetry). At 
absolute zero, the confinement/deconfinement transition - again, occurring when
the single-chain Mott gap is of the order of the Zeeman-like band splitting
energy - is a simple crossover\cite{Giamcom}.
This statement i.e the absence of a metallic phase at half-filling, can be
generalized for N-leg Hubbard ladders with N=3,4... (N is not too large).

First, the phase of perfect confinement is always equivalent to a N-leg 
Heisenberg ladder. Taking into account spinon-pair tunneling process between 
{\it successive} chains, it is indeed sufficient to induce
a pure spin-ladder with open boundaries in
the transverse direction. Based on known results on coupled spin chains, we
predict an insulating C0S0 phase for N {\it even}, and C0S1 for N 
{\it odd}\cite{Heinz}. 

Second, similar conclusions can be reached for the N-leg ladder 
far in the deconfined regime ($t_{\perp}$ large), analyzing the resulting
N-band model\cite{gene}. Let us briefly insist on the main arguments. At
half-filling, one gets the following Fermi velocities:
\begin{equation}
v_{j}=v_{\bar{\jmath}}=2t\hbox{\Large{\{}}1-
{\hbox{\large{(}}\frac{t_{\perp}}{t}\hbox{\large{)}}}^2
{\cos}^2\hbox{\large{(}}\frac{\pi j}{N+1}\hbox{\large{)}}
\hbox{\Large{\}}}^{1/2},
\end{equation}
with $\bar{\jmath}=N+1-j$. This results in:
\begin{equation}
v_1=v_N<v_2=v_{N-1}<... .
\end{equation}
Integrating properly 
the numerous RG-equations, one finds then a {\it decoupling} into
band pairs $(j,\bar{\jmath})$ and a hierarchy of energy scales
\begin{equation}
M_j\approx \exp-\pi v_j/U,
\end{equation}
where the band pairs scale successively towards the D-Mott state of the 
two-leg Hubbard ladder. For N {\it even}, all
excitations are then gapped (the phase is C0S0). For N {\it odd}, the remaining
band behaves like a single chain at commensurate filling resulting in a
C0S1 phase. As long as N is not too large, then the deconfined region is 
still insulating and there is no fundamental difference with the 
confined part, 
where in contrast $t_{\perp}$ is strongly 
suppressed: The deconfinement/confinement transition is still a crossover. No
metallic phase arises for large $t_{\perp}$ in the non-doped case.

This completely leaves open how the crossover should evolve for an 
an increasing number of coupled chains to give back the metal/insulating
transition observed in Bechgaard materials (N is very large). In particular, 
the decoupling of band pairs breaks down for 
large N $(v_1\approx v_2...)$ and the analysis of the  
RG-flow becomes very subtle mainly due to the plethora
of relevant interband coupling channels\cite{gene}. The fixed point for quite
large $t_{\perp}$ is not yet known.
A simplified (and much less rigorous) route to describe the 
strange metallic state of the TMTSF salt would be to start from the confined 
picture (with explicit spin-charge separation) and then to interpret
the relevant hopping as an induced ``self-doping'' on the 
chains, that are not obligatory equally doped\cite{Giam1}. 
Self-doping would appear as a possible
way to mimic the small deviation of the commensurate filling due to
the warping of the Fermi surface perpendicular to chain direction. 

Since here chains are not equally self-doped, then one does not predict any 
hole-pairing
effect: Charge excitations would be then those of the single-chain problem
i.e. doubly and empty occupied sites. For {\it small} induced self-doping i.e.
for $t_{\perp}$ close to m, the resulting
system might have similitudes with the lightly 
doped Hubbard chain\cite{Heinz,Giam1}. This simple picture
could explain the very small spectral weight $\delta(w)$
(Most of spectral weight comes from charge-gapped excitations) and the
unusual frequency-dependent conductivity in the finite frequency 
regime\cite{Controzzi}, observed on the basis of optical 
measurements\cite{Vescoli}.
\vskip 0.17cm
On the other hand, it is
important to remind that the TMTSF-family yields a good $T^2$-resistivity
showing the relatively {\it large} importance of the transverse hopping in this
system. The induced self-doping would not be so light in these materials
and then the 1D LL-prescription based on a single doped Hubbard chain 
naturally breaks down (We remind that LL behavior is rarely stable
in the N-chain model {\it away} from half-filling\cite{N-chain}). 
A possible way to reconcile Fermi-liquid behavior with optical 
datas would be to study the {\it effective} infinite-band model 
where bands with intermediate indices 
would be sufficiently doped whereas bands with small or large indices
would remain insulating. As soon as bands with intermediate indices are 
consequently doped, one indeed expects a weak-coupling
fixed point of Fermi-liquid type\cite{N-chain}.

\acknowledgments{We acknowledge discussions with L. Degiorgi, F.D.M. Haldane,
U. Ledermann, A. Nersesyan, R. Noack, T.M. Rice and thank
T. Giamarchi for a correspondence.}

\begin{appendix}

\section{Excitations at the Mott transition}

Note that the opening of the Mott gap produces
charge {\it fermionic kinks} confined to the chains. 

Using the transformations ({\it $K\approx 1$ denotes the bare LL exponent})
\begin{equation}
\Phi_{jc}=\sqrt{K}\tilde{\Phi}_{jc} \qquad 
\Theta_{jc}=\tilde{\Theta}_{jc}/\sqrt{K},
\end{equation}
the interacting part driven by umklapps in the chain j is of Sine-Gordon type:
\begin{equation}
{\cal H}_{int}=-g_u\cos\sqrt{4\pi K}\tilde{\Phi}_{jc}=
 -m \cos\sqrt{4\pi}\tilde{\Phi}_{jc}.
\end{equation}
Solutions of the underlying Sine-Gordon model are known as soliton-like; 
Solitons are fermionic particles:
\begin{equation}
{\cal F}^{Q_c^{\pm}}_{pj}=\kappa_{pj}\exp i\sqrt{\pi}Q_c^{\pm}(-p\tilde{\Phi}_{jc}+\tilde{\Theta}_{jc}),
\end{equation}
with the chosen gauge $\kappa_{+j}\kappa_{-j}=i$, and 
$Q_c^{\pm}=\pm 1$ refer to ``electron'' and ``hole'' like excitations
respectively. Therefore, it is not surprising that 
the charge part of the Hamiltonian density can be 
refermionized, as\cite{Giam1}
\begin{equation}
{\cal H}^c=\sum_j\ \pm iu{\cal F}_{pj}^{+}\partial_x{\cal F}_{pj}^{-}
-im{\cal F}_{+j}^{+}{\cal F}_{-j}^{-}+{\mathrm H.c.}
\end{equation}
Excitations can be viewed as pairs of doubly/empty occupied sites; a Kink
${\cal F}_{pj}^{Q_c^{\pm}}$ carries momentum $(-pQ_c^{\pm})2k_F$.
It should be noticed that fermionic (exchange) statistics means precisely:
\begin{equation}
{\cal F}_{pj}^{\pm}(x){\cal F}_{p^{\prime}j}^{\pm}(y)={\cal
  F}_{p^{\prime}j}^{\pm}(y){\cal F}_{pj}^{\pm}(x)\exp -i\gamma sgn(x-y),
\end{equation}
with 
\begin{equation}
\gamma=\pi. 
\end{equation}
Using Fourier and Bogoliubov transformations, the band structure of these 
solitons reads
\begin{equation}
E(k)=\pm\sqrt{(uk)^2+m^2}.
\end{equation}
This results in a semi-conducting picture, where the charge gap is equal to 
(2m). At zero energy, there is no way to create pairs of doubly/empty 
occupied sites. The ground state yields a long-range $4k_F$ CDW.

To study pair-hopping processes, one can rewrite:
\begin{equation}
{\cal H}_{int} =-2g_u\cos\sqrt{4\pi}\Phi_c^+\cos\sqrt{4\pi}\Phi_c^-.
\end{equation}
At the Mott transition, the pinning of these terms renders one-point
correlation functions {\it constants}
\begin{equation}
<\cos\sqrt{4\pi}\Phi_c^+>\ =\ <\cos\sqrt{4\pi}\Phi_c^->\ \approx m.
\end{equation}
We use normal ordering of phase exponentials in Sine-Gordon models:
\begin{equation}
<\cos\beta\Phi>\ =m^{K{\beta}^2/4\pi}:\cos\beta\Phi:,
\end{equation}
and the fact that the pinning phenomenon produces:
\begin{equation}
:\cos\sqrt{4\pi}\Phi_c^+:\ =\ :\cos\sqrt{4\pi}\Phi_c^-:\ =1.
\end{equation}
Due to cluster decomposition principle, the associated
two-point correlation functions
become also constants at large distance.
Note that the one-point correlation functions of the dual fields are 
still equal to zero (the 2-point correlation functions decay exponentially).

At the Mott energy, spin excitations or
so-called {\it spinons} tend to be also
localized on the chains. This carries the spin of a physical electron
and then is described by the operator:
\begin{equation}
\label{spinons}
{\cal S}^{\pm}_{pj}=\exp i\sqrt{\frac{\pi}{2}}Q_s^{\pm}(-p{\Phi}_{js}+
\Theta_{js}),
\end{equation}
with, $S^z=Q_s^{\pm}/2=\pm 1/2$.
Such objects are known to be located at wave vectors $pk_F$,
and obey a semionic statistics with $\gamma=\pi/2$\cite{K.-V}. 
In absence of charge fluctuations and for vanishing
$t_{\perp}$, spin fluctuations
in each chain are produced only by spinon pairs\cite{Haldane2}. 
\vskip 0.2cm
As for the antiferromagnetic Heisenberg chain, it is appropriate
to define the staggered spin operator as:
\begin{equation}
{\bf m}_j={\cal
  S}_{-j}^{\mu}{\pmb{$\sigma$}}({\cal S}_{+j}^{\mu^{\prime}})^*+{\mathrm H.c.},
\end{equation}
resulting explicitly in:
\begin{equation}
\label{spinon}
{\bf m}_j\propto
\hbox{\Large{(}} \cos\sqrt{2\pi}\Theta_{js},-\sin\sqrt{2\pi}\Theta_{js},
\cos\sqrt{2\pi}\Phi_{js}\hbox{\Large{)}}.
\end{equation}
Such definition of ${\bf m}_j$ respects the total breakup of physical 
electrons below the Mott scale.
Now, it is maybe 
important to compute explicitly the product ${\bf m}_1\cdot{\bf m}_2$ (Consult
subsection IV-A). One gets:
\begin{eqnarray}
{\bf m}_1{\bf m}_2 = &2&\cos\sqrt{2\pi}(\Theta_{1s}-\Theta_{2s})\\ \nonumber
&+&2\cos\sqrt{2\pi}\Phi_{1s}\cos\sqrt{2\pi}\Phi_{2s}\\ \nonumber
= &2&\cos\sqrt{4\pi}\Theta_s^- +\cos\sqrt{4\pi}\Phi_s^+
+\cos\sqrt{4\pi}\Phi_s^-.
\end{eqnarray}
>From Abelian equalities:
\begin{eqnarray}
{\cal S}^{+}_{p1}{\cal S}^{-}_{p2} +{\mathrm H.c.}&=&
\cos\sqrt{\pi}\Theta_s^-\cos\sqrt{\pi}\Phi_s^-,\\ \nonumber
{\cal S}^{+}_{p1}{\cal S}^{-}_{-p2}+{\mathrm H.c.} &=&
\cos\sqrt{\pi}\Theta_s^-\cos\sqrt{\pi}\Phi_s^+,
\end{eqnarray}
this also results in:
\begin{equation}
{\bf m}_1\cdot{\bf m}_2 = \hbox{\Large{[}}\sum_{p,p^{\prime}} 
{\cal S}^{+}_{p1}{\cal S}^{-}_{p^{\prime}2}+
{\mathrm H.c.}\hbox{\Large{]}}^2.
\end{equation}
This gives an explicit relationship between antiferromagnetic interchain
coupling and spinon-pair hopping.

Now, let us comment on the amplitude of the spinon pair hopping(s) in the
entrance of the Mott transition i.e. $l\rightarrow l_c$ [See
Eq. (31) in the Section IV]. At very short distances $l\rightarrow 0$,
pair hoppings are neglectable. Under renormalization
of the short-distance cut-off, the associated amplitude is known to
obey\cite{K-M}:
\begin{equation}
\frac{d\hat{g}}{dl}=(1-K)\hat{g}+z(1/K-K),
\end{equation}
with the {\it bare} conditions 
for $l\rightarrow 0$: $\hat{g}(0)=0$, $z(0)={t_{\perp}}^2/{E_F}^2$ and $(1/K-K)\sim U/v_F$ for small U. $E_F$ is
typically the Fermi energy. See e.g. Ref.\cite{TNG} page 226.
For $l\leq l_c$, this can be simplified as:
\begin{equation}
\frac{d\hat{g}}{dl}\approx z(l)\times U/v_F.
\end{equation}
{\it The amplitude of the pair-hopping(s) becomes finite approaching the Mott
transition.}
Using Eq. (\ref{flow}), we find:
\begin{equation}
\hat{g}(l_c)\propto \int_0^{l_c} z(l)dl=\int _0^{l_c} dz(l)
\approx {t_{\perp}}^2/m^2.
\end{equation}
We have neglected $z(0)$ in front of 
$z(l\rightarrow l_c)={t_{\perp}(l\rightarrow l_c)}^2/{E_F}^2
={t_{\perp}}^2/m^2$ because $m\ll E_F$. Furthermore, averaging explicitly
on the charge sector, we get
that for $T\rightarrow m$ the effective amplitude of the spinon 
pair-hopping(s) reads:
\begin{equation}
g_i(l_c)=\hat{g}(l_c)<\cos\sqrt{4\pi}\Phi_c^+>\ \propto {t_{\perp}}^2/m.
\end{equation}
Below the Mott transition $(l>l_c)$, $t_{\perp}(l)$ and $K(l)$ are
both renormalized to {\it zero} [See Part III, Eqs. (11) and (12)]. 
Therefore, $g_i$ now obeys:
\begin{equation}
\frac{d g_i}{dl}=g_i,
\end{equation}
with the new bare condition $g_i(l_c)={t_{\perp}}^2/m$. Far in the Infra-Red,
the spinon pair-hopping(s) will
``diverge'', producing a spin gap of the order of 
$J_{\perp}={t_{\perp}}^2/m$ in the spectrum (which can be exactly obtained
via refermionization, see Part IV B).

\section{Charge excitations in a LL}

In a LL, the charge Hamiltonian is plasmon-like:
\begin{equation}
H_{o}^c=\frac{u}{2\pi}\int dx\ \frac{1}{K}{(\rho_{c}-\rho_o)}^2
+K{(\nabla\Theta_{c})}^2.
\end{equation}
$\partial_x\Phi_{c}=(\rho_{c}-\rho_o)$ measures
fluctuations of charge density, and $\nabla\Theta_{c}$ is the
conjugate momentum to $\Phi_{c}$. The equations of motion are usual
d'Alembert equations. 

It is appropriate to use the chiral decomposition:
\begin{equation}
\Theta_{\pm}=\Theta_c\mp\frac{\Phi_c}{K}\cdot
\end{equation}
Again, $p=\pm$ refers to the direction of propagation (right or left).
The associated Hamiltonians are defined as
\begin{equation}
H_{\pm}=\frac{u}{4\pi}\int dx\ (\nabla\Theta_{\pm})^2.
\end{equation}
The objects $\Theta_{\pm}$ are chiral i.e. they obey\cite{K.-V}
\begin{equation}
[\Theta_{\pm}(x),\mp \frac{K}{2}\partial_y\Theta_{\pm}(y)]=i\delta(x-y).
\end{equation}
Chiral vertex operators of the effective Gaussian model read
\begin{equation}
{\cal L}_{\pm}^{Q_c^{\pm}}=\exp i\sqrt{\frac{\pi}{2}}Q_c^{\pm}\Theta_{\pm}.
\end{equation}
They describe charge excitations of a LL. In particular, these have a charge
$Q_c^{\pm}=\pm(1)$ because
\begin{equation}
[\hat{Q}_c,{\cal L}_p^{Q_c^{\pm}}]=Q_c^{\pm}{\cal L}_p^{Q_c^{\pm}}.
\end{equation}
The charge operator is correctly normalized as:
\begin{equation}
\hat{Q}_c=\sqrt{\frac{2}{\pi}}\int dx\ \partial_x\Phi_c.
\end{equation}
Such objects correspond to anyonic excitations since they obey anyonic 
commutation relations with:
\begin{equation}
\gamma=\frac{\pi}{2K}\cdot
\end{equation}
Note that the Hubbard interaction between physical electrons produces
(at low-energy) a change in the statistics of ``holons''. For the free
electron gas, one gets $K=1$ and then holons are rather semions or
half-electrons.

\end{appendix}


\begin{references}
\bibitem{remark} The vanishing of spin backscattering in the case of two
degenerate bands can be understood as follows. Here, there are only two Fermi
points. Therefore, one cannot distinguish anymore between intra- and 
interband interactions.
Looking at equation flows of Ref.\cite{gene} and equating
$c_{11}^{\sigma}=c_{12}^{\sigma}=g_s$ one immediately obtains the 
expected (single-chain) result.
\bibitem{rem} One must keep the backward hopping although it is oscillating
with a prefactor $(-1)^j$, because it generates pair hoppings
at $q=0$.
\bibitem{rem0} The term $g_6$ has been omitted because it will give only
logarithmic corrections to the spin gap.
\bibitem{symmetry} The $SU(2)_2$ symmetry comes from the fact that the 
$SU(2)_2$
currents and the primary field with spin-1 
can be explicitly written in function of the Majorana fermions 
$\xi_i$. The ${\cal Z}_2$ (or Ising) symmetry results from the invariance
of the model under $\rho\rightarrow -\rho$. This decomposition is known as a
conformal embedding.
\bibitem{MI-tran} For the one-chain case, one gets: $[\hat{Q}_{jc},{\cal F}_{pj}^-]=-\sqrt{2K}{\cal F}_{pj}^-$ with $\hat{Q}_{jc}=\sqrt{2K/\pi}\int dx\ 
\partial_x
\tilde{\Phi}_{jc}$. Since physical objects are now spinless fermions with
charge $Q_c^-=-1$, then this results in $K\rightarrow 1/2$ near half-filling\cite{Schulz}.
\end{references}
\end{document}